\documentclass[
    sigconf
]{acmart}

\copyrightyear{2023}
\acmYear{2023}
\setcopyright{rightsretained}
\acmConference[RecSys '23]{Seventeenth ACM Conference on Recommender Systems}{September 18--22, 2023}{Singapore, Singapore}
\acmBooktitle{Seventeenth ACM Conference on Recommender Systems (RecSys '23), September 18--22, 2023, Singapore, Singapore}\acmDOI{10.1145/3604915.3608774}
\acmISBN{979-8-4007-0241-9/23/09}

\usepackage{amsthm}
\usepackage[ruled]{algorithm2e}
\usepackage{multirow}

\newcommand{\argmax}{\mathop{\rm arg~max}\limits}
\newcommand{\argmin}{\mathop{\rm arg~min}\limits}
\newcommand{\bE}{\mathbb{E}}
\newcommand{\bP}{\mathbb{P}}

\newcommand{\mC}{\mathcal{C}}
\newcommand{\mJ}{\mathcal{J}}
\newcommand{\rank}{\mathrm{rank}}
\newcommand{\rk}{\mathbf{rk}}
\newcommand{\sw}{\mathbf{SW}}
\newcommand{\usw}{\underline{\sw}}
\newcommand{\T}{\mathsf{T}}

\theoremstyle{definition}
\newtheorem{definition}{Definition}
\newtheorem{assumption}{Assumption}

\theoremstyle{plain}

\theoremstyle{remark}

\usepackage{booktabs}
\usepackage[colorinlistoftodos,prependcaption]{todonotes}
\renewcommand{\paragraph}[1]{%
\noindentparagraph{\textbf{\textup{#1.}}}
}%

\begin{document}
\title{Fast and Examination-agnostic Reciprocal Recommendation in Matching Markets}

\author{Yoji Tomita}
\affiliation{%
  \institution{CyberAgent, Inc.}
  \city{Tokyo}
  \country{Japan}
}
\email{tomita_yoji@cyberagent.co.jp}

\author{Riku Togashi}
\affiliation{%
  \institution{CyberAgent, Inc.}
  \city{Tokyo}
  \country{Japan}
}
\email{togashi_riku@cyberagent.co.jp}

\author{Yuriko Hashizume}
\affiliation{%
  \institution{CyberAgent, Inc.}
  \city{Tokyo}
  \country{Japan}
}
\email{hashizume_yuriko@cyberagent.co.jp}

\author{Naoto Ohsaka}
\affiliation{%
  \institution{CyberAgent, Inc.}
  \city{Tokyo}
  \country{Japan}
}
\email{ohsaka_naoto@cyberagent.co.jp}


\begin{abstract}
In matching markets such as job posting and online dating platforms, the recommender system plays a critical role in the success of the platform. Unlike standard recommender systems that suggest items to users, reciprocal recommender systems (RRSs) that suggest other users must take into account the mutual interests of users. In addition, ensuring that recommendation opportunities do not disproportionately favor popular users is essential for the total number of matches and for fairness among users. Existing recommendation methods in matching markets, however, face computational challenges on real-world scale platforms and depend on specific examination functions in the position-based model (PBM). In this paper, we introduce the reciprocal recommendation method based on the matching with transferable utility (TU matching) model in the context of ranking recommendations in matching markets, and propose a faster and examination-agnostic algorithm. Furthermore, we evaluate our approach on experiments with synthetic data and real-world data from an online dating platform in Japan. Our method performs better than or as well as existing methods in terms of the total number of matches and works well even in relatively large datasets for which one existing method does not work.
\end{abstract}



\maketitle

\section{Introduction}
\label{sec:introduction}
In matching markets, such as job posting and online dating platforms, reciprocal recommender systems (RRSs) play a crucial role in the success of the platforms~\cite{pizzato2010recon,mine2013reciprocal}.
Unlike conventional recommender systems that suggest items to users, RRSs that suggest other users need to take into account two aspects that do not appear in conventional recommender systems.
First, recommendation results need to be based on the mutual preferences of users.
In job posting platforms, for example, even if one job seeker has a strong interest in one job post, the employer might not be interested at all in the job candidate;
the ``match'' between the job seeker and the employer will not be established in this case.
Therefore, recommending job posts to a job seeker only according to the seeker's interest is not effective,
and hence RRSs should recommend only when both users are mutually interested.
Second, recommendation opportunities should not concentrate on a few popular users.
The number of matches that one user can make is often physically limited because it is impossible for one employer to have interviews with all candidates.
Thus, distributing match opportunities widely among users is important to maximize the total number of matches, and fairness among users is important in matching markets.

To take into account the first aspect (i.e., the mutuality of preferences), most reciprocal recommendation methods aggregate two unilateral preference scores (of job candidates toward employers and vice versa) into a bilateral preference score via some aggregate functions.
Unilateral preference scores can be generally calculated by standard recommendation methods such as matrix factorization~\cite{hu2008collaborative,koren2009matrix}.
Arithmetic mean, geometric mean, or harmonic mean are typically used as the aggregate function in the literature and in practice.
However, most of the aggregate functions do not reflect the second aspect, the necessity of avoiding concentration on a few popular users, and this issue has not been analyzed enough in the literature.

To tackle this issue, \citet{Su2022-hx} recently consider the problem of ranking recommendation in the matching markets, including job posting platforms.
They define the social welfare function as the expected total number of matches and propose a method to choose a stochastic ranking recommendation policy that maximizes the approximate social welfare function.
In the experiment simulating a matching market, their method substantially improves the total number of matches compared to both the naive method based on naive unilateral preferences and the reciprocal method based on the product of two unilateral preferences.
However, there are still two practical challenges to using \citet{Su2022-hx}'s method in real-world platforms.
First, their method involves a doubly stochastic matrix, whose size is quadratic in the number of users in the reactive side (e.g., employers), for each proactive side user (e.g., job candidates) as variables, and solves a convex optimization problem with constraints.
In real-world scale matching platforms, it is very expensive to directly solve the convex optimization problem of such sizes in terms of both space and time complexity.
In addition, implementing the stochastic ranking recommendation is challenging because the Birkhoff--von-Neumann decomposition~\cite{birkhoff1946three} of doubly stochastic matrices for each proactive user is also computationally prohibitive.
Second, their method depends on the specific form of the examination functions in the position-based model (PBM).
In PBM, the examination function maps a ranking that an item listed on to a probability with which a user has attention to the item.
Although some methods to estimate the examination function are proposed, it is still hard in practice to estimate it correctly because the examination functions are specific to each application and may be different among individual users.
\citet{Su2022-hx} use an examination function for both the objective function to optimize in their method and the evaluation via Monte Carlo simulations, and they do not consider the problem of misspecification of examination functions.

In the literature on the matching theory~\cite{gale1962college,roth1992two}, economists and computer scientists have developed various matching algorithms that seek to match two sides of agents appropriately according to the mutual preferences of agents and their matching capacity, which correspond to two aspects of RRSs discussed above. 
Especially, matching with transferable utility (TU matching) models~\cite{shapley1971assignment} have been developed to analyze the job markets~\cite{kelso1982job} and the marriage markets~\cite{becker1973theory,becker1974theory,Choo2006-dd}.
In these models, monetary transfers are assumed to occur between two agents matched, and the amounts of transfers are adjusted so as to balance the demands from both sides in the matching market, just as prices in the standard economic markets.
The resulting matching, which is called ``equilibrium matching'' in this paper, reflects the mutuality of preferences and avoids the concentration of matching on a few popular people, and thus it would be effective to apply it in the reciprocal recommendations.

In this paper, we consider the reciprocal recommendation method based on the TU matching model of \citet{Choo2006-dd}, 
in the setting of the ranking recommendation in the matching markets. 
Our proposed \emph{TU} method computes a deterministic ranking policy more efficiently in terms of both space and time complexity than \citet{Su2022-hx}'s method, and it is independent of a form of the examination function.

The contribution of this paper is as follows.
First, in the setting of ranking recommendation in the matching markets, we propose a reciprocal recommendation method based on TU matching model and develop an efficient algorithm.
Second, we evaluate our \emph{TU} method in experiments with synthetic data and real-world data from an online dating platform in Japan, comparing with baselines including \emph{SW} method of \citet{Su2022-hx}.
Our \emph{TU} method performs at least as well as \emph{SW} method in terms of the expected total number of matches, and outperforms the conventional RRS methods such as \emph{Naive} and \emph{Reciprocal} methods.
Moreover, \emph{TU} method outperforms \emph{SW} with the misspecified examination function in most cases, and it also works well even in relatively large datasets where \emph{SW} method does not work.

\subsection{Related work}
\label{subsec:related-works}
\paragraph{Reciprocal recommendation}
Of particular distinction in reciprocal recommendation from the standard item recommendation is that a system must consider the bilateral preferences of proactive and reactive users.
Most of the conventional methods estimate the preferences of each side of users by using a corresponding model and then use the bilateral preference estimates to predict a reciprocal preference.
As the first step of unilateral preferences can be solved by standard recommender methods,
modeling reciprocal preferences is often reduced into the aggregation of bilateral preferences.
In this context, various aggregation functions have been proposed, e.g., arithmetic mean~\cite{neve2019latent,neve2019aggregation}, geometric mean~\cite{neve2019latent,neve2019aggregation}, harmonic mean~\cite{pizzato2010recon,xia2015reciprocal}, cross-ratio uniform~\cite{appel2017cross}, matrix multiplication~\cite{jacobsen2019s}, the multiplicative inverse of rank multiplication~\cite{mine2013reciprocal}, and weighted mean with optimized weighting parameters~\cite{kleinerman2018optimally}.
\citet{neve2019latent} proposed an early method that optimizes two independent matrix factorization models of unilateral preferences and then aggregates two preference estimates using arithmetic, geometric, or harmonic means.
However, such aggregation functions highly depend on the scale and domain of unilateral preference estimates and lack a theoretical foundation.
By contrast, recent fairness-aware recommendation 
do not consider such aggregation functions by directly maximizing a welfare function~\cite{do2021two,do2022optimizing}.

\paragraph{Inefficiency issues in post-processing approach based on doubly stochastic matrices}
Fairness-aware methods based on the post-processing approach often rely on the post-estimation of doubly stochastic matrices, which represent stochastic recommendation policies for users~\cite{singh2018fairness,do2021two,usunier2022fast,do2022optimizing,saito2022fair,Su2022-hx}.
Methods in this category may be impractical due to the quadratic computational cost with respect to the number of items for each user.
Recent methods alleviate this issue by using Frank-Wolfe algorithms that enable efficient parameter updates in the quasi-linear cost with respect to the number of items for each user~\cite{do2021two,do2022optimizing}.
The practical implementation of probabilistic ranking systems is also challenging because the pre-computation step for sampling a ranked list from a doubly stochastic matrix can be costly due to the Birkhoff--von-Neumann decomposition~\cite{birkhoff1946three}, which involves iteratively finding a perfect matching on the matrix.
Moreover, although the main interest of conventional studies is the fairness guarantee for algorithms, most of the provable guarantee does not hold in practice as the availability of the true preferences and true examination probabilities is often assumed.

\paragraph{Matching theory and reciprocal recommendations based on matching theory}
Our methodology builds upon the literature of matching theory~\cite{roth1992two,manlove2013algorithmics} originated by \citet{gale1962college}, which has been widely applied to a variety of applications, including hospital-resident assignment~\cite{roth1999redesign,kamada2015efficient}, school matching~\cite{abdulkadirouglu2005boston,abdulkadirouglu2005college,abdulkadirouglu2005new}, and paper-reviewer assignment~\cite{lian2018conference,garg2010assigning,ivan2021peer}.
Especially, the matching with transferable utility model (TU model)~\cite{shapley1971assignment,chiappori2017matching} considers matching markets where agents matched can transfer utility or money between them. Economists have applied the TU matching models to analyze the job market~\cite{kelso1982job} and the marriage market~\cite{becker1973theory,becker1974theory}.
\citet{Choo2006-dd} develops an empirical model of the marriage market and applies it to the empirical study of the real marriage market in the US, and several works have developed its generalized model and estimation algorithms~\cite{decker2013unique,Galichon2022-dn}.

Although there are a few studies that apply the matching theory literature to the problem of reciprocal recommendation~\cite{saini2019privatejobmatch,eskandanian2020using,pizzato2011stochastic}, the interaction of reciprocal recommendation and matching theory has been relatively less analyzed.
The most similar work to this paper in terms of methodology is \citet{Chen2023-si}, which considers recommending users based on the equilibrium matching of \citet{Choo2006-dd} model in online dating.
However, the main method that they use in the online experiment separates users into several groups and estimates preferences for each group, and thus it does not correspond directly to the fully personalized recommendation\footnote{\citet{Chen2023-si} also considers personalized recommendation that built on preferences estimated via MF in offline experiments in the appendix of their paper, but their experimental setting is limited and its impact on the total number of matches is not analyzed enough.}.
In addition, the main issue that they consider is the inequality of matching opportunities among users rather than the total number of matches.
\citet{Tomita2022-yv} also discuss the reciprocal recommendation method based on \citet{Choo2006-dd} model, but they did not describe the whole algorithm of the method and did not evaluate the performance in any experiments.
In this paper, we propose a fully-personalized reciprocal recommendation method based on TU matching of \citet{Choo2006-dd} model in the framework of ranking recommendation in matching markets~\cite{Su2022-hx}, and evaluate its performance in terms of the total number of matches in experiments with synthetic data and real-world data.

\begin{table*}[t]
    \centering
    \begin{tabular}{c|l}
        \toprule
        \textbf{notation} & \textbf{description} \\
        \midrule
        $\mC$ & set of job candidates  \\
        $\mJ$ & set of job posts \\
        $p_{c,j}$ & estimated preference score of candidate $c \in \mC$ to employer $j \in \mJ$ \\
        $p_{j,c}$ & estimated preference score of employer $j \in \mJ$ to candidate $c \in \mC$ \\
        $\Sigma_{|\mJ|}$ & set of all possible ranked lists of employers $\mJ$ \\
        $\sigma \colon \mC \to \Sigma_{|\mJ|}$ &
            a deterministic ranking policy of each job candidate $c \in \mC$
        \\
        $\pi \colon \mC \to \Delta(\Sigma_{|\mJ|})$ &
            a stochastic ranking policy is a mapping from a candidate
        \\
        $v \colon \mathbb{N} \to [0,1]$ & examination function, e.g., $v(x) = 1/x$ or $1/\log(1+x)$ \\
        $\rank(j \mid \sigma(c) )$ &
            employer $j$'s rank in the ranked list $\sigma(c)$ \\
        $\rk^\pi_j(c)$ &
            candidate $c$'s rank in employer $j$'s list (when $c$ applied to $j$) \\
        $\bP_{c,j}^\pi$ & probability that candidate $c \in \mC$ applies to employer $j \in \mJ$ \\
        $\bP_{j,c}^\pi$ & probability that employer $j \in \mJ$ matches to candidate $c \in \mC$ \\
        $\sw(\pi)$ & social welfare of stochastic ranking policy $\pi$ \\
        $M_c^\pi$ & a doubly stochastic matrix with size of $|\mJ|\times |\mJ|$ for $c$ reduced from $\pi$\\
        \bottomrule
    \end{tabular}
    \caption{Notations and definitions frequently used in this paper.}
    \label{tab:notations}
\end{table*}

\section{Setup}
\label{sec:setup}
In this section, we introduce a framework for ranking recommendations in matching markets according to \citet{Su2022-hx}, and several re-ranking methods such as the naive, reciprocal, and social welfare maximization \cite{Su2022-hx}.
Table~\ref{tab:notations} lists the notations frequently used in this paper.

\subsection{Framework for recommendation in matching markets}
\label{subsec:framework}

Following \citet{Su2022-hx}, we describe the framework of matching markets in the context of job posting platforms.
This framework can be applied to other matching markets such as online dating platforms.
In this market, job candidates (proactive side) first browse a list of ranked job posts/employers and apply to the ones they are interested in. The employer (reactive side) receives a list of job candidates who applied to them and then selects the ones he/she considers qualified for the job. 

We now present a formal model of this market.
Let $\mC$ be the set of job candidates, and $\mJ$ be the set of job posts or employers.
We note that both sets are finite ($|\mC|, |\mJ| < +\infty$).
Each candidate $c \in \mC$ has a preference over the set of job posts $\mJ$, and  each employer $j \in \mJ$ has a preference over the candidate set $\mC$.

Although the platform does not know the true preferences of job candidates and employers, we assume that their estimations are available by using some recommendation methods.
Let $p_{c,j} \in [0,1]$ be the estimated preference  of candidate $c \in \mC $ to employer $j \in \mJ$, and $p_{j, c} \in [0,1]$ be that of $j$ to $c$.
The score $p_{c,j}$ can be thought of as the probability with which $c$ has relevance with $j$ conditioning that $c$ examined $j$;
a similar analogy can be applied to $p_{j,c}$.

The platform can choose a ranked list of employers sent to each job candidate.
Let $\Sigma_{|\mJ|}$ be the set of all possible ranked lists of employers $\mJ$.
A deterministic ranking policy $\sigma: \mC \to \Sigma_{|\mJ|}$ is a mapping from each job candidate $c \in \mC$ to a ranked list of employers $\sigma(c) \in \Sigma_{|\mJ|}$ that is sent to the candidate $c$.
A stochastic ranking policy $\pi : \mC \to \Delta(\Sigma_{|\mJ|})$ is a mapping from a candidate $c \in \mC$ to a probability distribution of ranked lists of employers $\pi(\cdot \mid c) \in \Delta(\Sigma_{|\mJ|})$.
In this work, we study how to design deterministic ranking policies (the naive, reciprocal, and our methods mentioned below), while the focus of \citet{Su2022-hx} is on stochastic ranking policies. 
We also note that a deterministic ranking policy $\sigma$ is a special case of stochastic ranking policies that take ranking $\sigma(c)$ for each candidate $c \in \mC$ with probability $1$, and thus we explain the framework and objective in terms of stochastic policies for simplicity below.

After obtaining a ranked list of employers $\sigma(c)$, each candidate $c \in \mC$ chooses whether to apply for each employer.
In this framework, we use the position-based model (PBM) \cite{Joachims2017-ix}.
Conditioning on the ranked list $\sigma(c)$, the probability with which the candidate $c$  applies to an employer $j$ is
\begin{equation}
    \bP\left(c \text{ applies to } j \mid \sigma(c)\right) = v(\rank(j \mid \sigma(c))) \cdot p_{c,j}, \notag
\end{equation}
where $\rank(j \mid \sigma(c) ) \in \{1,2,\dots,|\mJ|\}$ is the rank of the employer $j$ in the ranked list $\sigma(c)$, and $v$ is an examination function, that is, $v(k) \in [0,1]$ is a probability with which the candidate examines the employer positioning on rank $k$.
Typically, $v(x) = 1/x$ or $1/\log(1+x)$ are used. 
With a stochastic ranking policy $\pi$, the probability with which a candidate $c \in \mC$ applies to an employer $j \in \mJ$, denoted by $\bP_{c,j}^\pi$ can be calculated as below:
\begin{align}
     \bP_{c,j}^\pi &= \bP\left( c \text{ applies to } j \mid \pi \right) \notag \\
    &= \sum_{\sigma(c) \in \Sigma_{|\mJ|}} \pi(\sigma(c) \mid c) \cdot \bP(c \text{ applies to } j \mid \sigma(c))\notag \\
    &= \sum_{\sigma(c) \in \Sigma_{|\mJ|}} \pi(\sigma(c) \mid c) \cdot v(\rank( j \mid \sigma(c))) \cdot p_{c,j}. \label{eq:p_cj^pi}
\end{align}

After the application process by the job candidates, each employer $j \in \mJ$ receives the list of candidates who applied to the employer $j$.
The list is sorted in descending order according to the employer's preference $p_{j,\cdot}$, beforehand by the platform.
Employers examine the lists and then choose candidates to ``match'', that is, to get in touch with or to conduct an interview with.
As in the case of job candidates, we also use the PBM for this employer's process.
The probability with which the employer $j \in \mJ$ matches to $c \in \mC$ conditioning on that $c$ applied to $j$, denoted by $\bP_{j,c}^\pi$ is
\begin{align}
    \bP_{j,c}^\pi &= \bP\left( j \text{ matches to } c \mid c \text{ applied to } j,\ \pi\right) \notag\\
    &=  \bE\left[v\left(\rk^\pi_{j}(c)\right) \right] \cdot p_{j,c} \notag
\end{align}
where $\rk^\pi_j(c)$ is a rank of $c$ in the list for the employer $j$ conditioning on the policy $\pi$ and that $c$ applied to $j$.
We note that $\rk^\pi_j(c)$ is a random variable even if $\pi$ is deterministic because it relies on the set of candidates who also applied for the employer $j$.

Given a policy $\pi$, the \emph{social welfare} is defined as the expected total number of matches~\cite{Su2022-hx}:
\begin{equation}\label{eq:sw}
    \sw(\pi) \coloneqq \sum_{c \in \mC}\sum_{j \in \mJ} \bP^\pi_{c,j} \bP_{j,c}^\pi.
\end{equation}
The platform's objective is to find an appropriate policy $\pi$ such that the social welfare $\sw(\pi)$ is maximized.

\subsection{Existing methods}
\label{subsec:existing-methods}
Given the matching market explained above, we highlight popular existing methods, the naive methods, reciprocal methods, and the approximate social welfare maximization approach \cite{Su2022-hx}.

\paragraph{Naive method}
One of the most popular methods in practice is to re-rank employers naively according to job candidates' preferences.
For each job candidate $c \in \mC$, let $\sigma^{\mathrm{naive}}(c)$ be the ranked list of employers $\mJ$, which is sorted in descending order according to the candidate $c$'s preference score $p_{c, \cdot}$.

\paragraph{Reciprocal method}
Many works in the RRSs use aggregation functions that aggregate two unilateral scores $p_{c,j}, p_{j,c}$ into a reciprocal score $p^\mathrm{rec}_{c,j}$ to take into account the mutuality of preferences.
One of the most popular aggregation functions is the geometric mean, or equivalently in terms of the order, the product of job candidates' and employers' preference scores $p^{\mathrm{rec}}_{c,j} = p_{c,j} \cdot p_{j,c}$~\cite{neve2019aggregation,neve2019latent}.
In the reciprocal method, the platform provides the job candidate $c$ with the ranked list of employers $\sigma^{\mathrm{rec}}(c) \in \Sigma_{|\mJ|}$ which is sorted in descending order according to the reciprocal scores $p^{\mathrm{rec}}_{c,\cdot}$.

\begin{algorithm*}[t]
    \caption{Social-Welfare Optimization via Frank-Wolfe \cite{Su2022-hx}}\label{alg:sw}
    \SetKw{initialize}{initialize}
    \SetKw{return}{return}
    \KwIn{preference matrices $(p_{c,j})_{(c,j)\in \mC\times \mJ}$ and  $(p_{j,c})_{(j,c)\in \mJ\times \mC}$, examination function $v(\cdot)$, timesteps $T$, learning rate $\eta_t$}
    \KwOut{the doubly stochastic matrices $M_\mC^\pi$}
    \initialize{$M_c^\pi \leftarrow \mathbf{1}\mathbf{1}^\T / |\mJ|$, $\forall c \in \mC$}\;
    \For{$t = 1,\dots T$}{
        $S^*_{\mC} \in \argmin_{S_\mC} -\nabla\usw(M_\mC^\pi)^\T S_\mC $ s.t. $\mathbf{1}^\T S_c = \mathbf{1}^\T, \ S_c \mathbf{1} = \mathbf{1},\ \forall c \in \mC,\ S_\mC = (S_c)_{c \in \mC}$\;
        $M_\mC^\pi \leftarrow (1 - \eta_t)M_{\mC}^\pi + \eta_t S^*_{\mC}$\;
    }
    \return{$M_{\mC}^\pi$}\;
\end{algorithm*}

\paragraph{Approximate social welfare maximization \cite{Su2022-hx}}
Unlike the naive and reciprocal methods, the approximate social welfare maximization method \cite{Su2022-hx} optimizes a stochastic policy directly. The probability~\eqref{eq:p_cj^pi} with which candidate $c \in \mC$ applies to employer $j \in \mJ$ in a stochastic policy $\pi$ can be expressed as:
\begin{align}
    \bP_{c,j}^\pi \notag
    &= \sum_{\sigma(c) \in \Sigma_{|\mJ|}} \pi(\sigma(c) \mid c) \cdot v(\rank( j \mid \sigma(c))) \cdot p_{c,j}\notag\\
    &= p_{c,j} \sum_{k = 1}^{|\mJ|} \bP_{\sigma(c) \sim \pi(\cdot \mid c)}\Big(\rank\left(j \mid \sigma(c)\right) = k\Big)\cdot v(k)\notag\\
    &= p_{c,j} \sum_{k = 1}^{|\mJ|} M^\pi_{c}(j,k)\cdot v(k) \label{eq:p_cj^pi-dsm}
\end{align}
where $M_c^\pi \in [0,1]^{|\mJ|\times |\mJ|}$ is the doubly stochastic matrix reduced from $\pi(\cdot \mid c)$, that is, $M_c^\pi(j, k)$ is the probability with which the employer $j$ is ranked at position $k$ for the candidate $c$ in the stochastic policy $\pi$.
In the PBM, the probability with which $c$ applies to $j$ relies only on the position $k$ where the employer $j$ is displayed and preference score $p_{c,j}$, and not on other employers' position.
Therefore, the application probability~\eqref{eq:p_cj^pi-dsm} can be expressed in terms of the doubly stochastic matrix $M_c^\pi$ reduced from the stochastic policy $\pi(\cdot \mid c)$.

Then the social welfare~\eqref{eq:sw} is:
\begin{align}
    \sw(\pi) &= \sum_{c \in \mC}\sum_{j \in \mJ} \bP^\pi_{c,j} \bP_{j,c}^\pi \notag\\
    &= \sum_{c \in \mC} \sum_{j \in \mJ} p_{c,j} \cdot p_{j,c} \cdot \bE\left[v\left(\rk^\pi_{j}(c)\right) \right]  \sum_{k = 1}^{|\mJ|} M^\pi_{c}(j,k)\cdot v(k). \label{eq:sw2}
\end{align}
However, the social welfare~\eqref{eq:sw2} cannot be directly optimized because the rank of $c$ in $j$'s list $\rk^\pi_{j}(c)$ depends on other candidates' applications and thus the term $\bE\left[v\left(\rk^\pi_{j}(c)\right) \right]$ is too complex to tackle with.
\citet{Su2022-hx} derive a lower bounds $\underline{\sw}\left(M_{\mC}^\pi\right)$ of $\sw(\pi)$ under the assumption of convexity of examination function $v$: 
\begin{align}
    \usw \left(M_{\mC}^\pi\right) \coloneqq &\sum_{c \in \mC}\sum_{j \in \mJ} p_{c,j}\cdot p_{j,c} \notag \\
    &\cdot v\left(1 + \sum_{c': p_{j,c'} > p_{j,c}}p_{c',j}\sum_{k=1}^{|\mJ|}M_{c'}^{\pi}(j,k)\cdot v(k)\right) \notag \\
    &\cdot \sum_{k=1}^{|\mJ|}M_{c}^\pi(j,k)\cdot v(k), \label{eq:approximate-sw}
\end{align}
where $M_\mC^\pi \coloneqq (M_c^\pi)_{c \in \mC}$. They propose the \emph{SW} method that optimizes this approximated social welfare instead of $\sw(\pi)$:
\begin{gather}
    \label{eq:opt-sw}
    \max_{M^\pi_\mC} \ \usw\left( M_\mC^\pi \right), \quad
    \mathrm{s.t.} \ \mathbf{1}^\T M_c^\pi = \mathbf{1}^\T, \ M_c^\pi \mathbf{1} = \mathbf{1}, \ \forall c \in \mC.
\end{gather}

Several approaches can be used to solve the convex optimization problem~\eqref{eq:opt-sw}.
\citet{Su2022-hx} use the Frank-Wolfe approach, which is summarized as Algorithm~\ref{alg:sw}.

However, there are some practical difficulties in implementing this social welfare optimization approach in real-world matching markets as discussed in Section~\ref{sec:introduction}.
The objective function~\eqref{eq:approximate-sw} includes the true examination function $v$, whose estimation is hard in practice.
Moreover, materializing the doubly stochastic matrices for all candidates is unrealistic due to its computational cost of $O(|\mC||\mJ|^2)$ for each timestep, and the precomputation for efficient sampling of rankings from a doubly stochastic matrix involves Birkhoff--von-Neumann decomposition, which is very costly to solve for all job candidates; we further discuss this point in Section~\ref{subsec:inference}.

\section{TU matching approach}
In this section, we propose a reciprocal recommendation algorithm based on the matching with transferable utility (TU matching) model.
In Section~\ref{subsec:TU}, we introduce \citet{Choo2006-dd}'s model of TU matching and explain the concrete algorithm.
In addition, we develop a practical inference implementation using efficient vector search algorithms \cite{shrivastava2014asymmetric} in Section~\ref{subsec:inference}, which is contrast to existing stochastic methods that need pre-computation of Birkhoff-von Neumann decomposition in inference time.

\subsection{TU matching model}
\label{subsec:TU}

In the model discussed below, we interpret that estimated preference scores $p_{c,j}$ and $p_{j,c}$ are the estimation of surplus or gains that they obtain when they match with each other.
When the candidate $c \in \mC$ and the employer $j \in \mJ$ match, they split surplus
\begin{equation}
    p_{c,j} +\epsilon_{c,j} + p_{j,c} + \epsilon_{j,c} \notag
\end{equation}
where $\epsilon_{c,j}$ and $\epsilon_{j,c}$ are the estimation error of preferences $p_{c,j}$, $p_{j,c}$ respectively, which are assumed to be identically and independently distributed from some probability distribution $F, G$.
In the TU matching model, we assume that monetary transfer occurs between a job candidate and an employer when they are matched.
Transfers are interpreted in physical markets as, for example, wages in the job market, household sharing in the marriage market, or gifts in dating markets.
Although there may not be such explicit transfers between users on online platforms, we assume that some implicit transfers occur and use virtual transfers as a tool to derive an appropriate matching.
Let $\tau_{c,j}$ be the transfer amount that the employer $j$ gives to $c$ when they match.
In this case, the candidate $c$/the employer $j$ obtains
\begin{equation}
    p_{c,j} + \epsilon_{c,j} + \tau_{c,j}, \qquad
    p_{j,c} + \epsilon_{j,c} - \tau_{c,j} \notag
\end{equation}
from the matching, respectively.
In this matching market, each agent chooses one that maximizes his/her gain from the set of opponents.
They also have an outside option, or remaining ``unmatched'' in this market, which is denoted as ``$0$''.
The stochastic demand of the candidate $c \in \mC$ toward the employer $j \in \mJ \cup \{0\}$, that is, the probability with which $c$ chooses $j$ is
\begin{equation}
    \mu_{c,j} = \bP_{\epsilon_{c,\cdot}}\left( j \in \argmax_{j' \in \mJ\cup\{0\}}\left(p_{c,j'} + \epsilon_{c,j'} + \tau_{c,j'}\right)\right), \label{eq:cand-demand}
\end{equation}
and the stochastic demand of the employer $j \in \mJ$ toward $c \in \mC\cup\{0\}$ (the probability with which $j$ chooses $c$) is
\begin{equation}
    \mu_{j,c} = \bP_{\epsilon_{j,\cdot}}\left( c \in \argmax_{c' \in \mC\cup\{0\}}\left(p_{j,c'} + \epsilon_{j,c'} - \tau_{c',j}\right) \right), \label{eq:employer-demand}
\end{equation}
where we assume $p_{c,0} = p_{0,j} = \tau_{c,0} = \tau_{0,j} = 0$, $\epsilon_{c,0} \sim F$ and $\epsilon_{j,0} \sim G$ i.i.d. for any $c \in \mC, j \in \mJ$.\footnote{The existence of an outside option is not a crucial assumption because the model and method can be written and derived without the outside option ``$0$''.
We contain $0$ following the tradition of matching with transferable utility literature \cite{Choo2006-dd} and for the simplicity of notation and method.
}

The equilibrium is a situation where both sides of stochastic demands $\mu_{c,j}, \mu_{j,c}$ coincide by the transfers' adjustment, just as the market price in the standard physical market, which adjusts demands and supplies.
We call the adjusted transfers $\tau^* = (\tau^*_{c,j})_{(c,j) \in \mC\times\mJ}$ and (fractional) matching $\mu^* = (\mu^*_{c,j})_{(c,j) \in \mC \times \mJ}$ as equilibrium matching.

\begin{definition}
An equilibrium matching $(\mu^*, \tau^*)$ satisfies the following conditions:
\begin{enumerate}
    \item $\mu^*$ satisfies
    \begin{align}
        &\mu_{c,0}^* + \sum_{j \in \mJ}\mu^*_{c,j} = 1 \qquad \forall c \in \mC, \label{eq:constraint1} \\
        &\mu_{0,j}^* + \sum_{c \in \mC}\mu^*_{c,j} = 1 \qquad \forall j \in \mJ. \label{eq:constraint2} 
    \end{align}
    \item $\mu^*$ equals both sides of demands:
    \begin{align}
        \mu^*_{c,j} &= \bP_{\epsilon_{c,\cdot}}\left( j \in \argmax_{j' \in \mJ\cup\{0\}}\left(p_{c,j'} + \epsilon_{c,j'} + \tau^*_{c,j'}\right)\right) \notag\\
        &= \bP_{\epsilon_{j,\cdot}}\left( c \in \argmax_{c' \in \mC\cup\{0\}}\left(p_{j,c'} + \epsilon_{j,c'} - \tau^*_{c',j}\right) \right)\notag
    \end{align}
    for all $(c,j) \in \mC \times \mJ$.
\end{enumerate}
\end{definition}

The intuition of the equilibrium matching in TU matching model is as follows.
For a popular candidate $c$, $p_{j,c}$ is generally high for many employers, and the demands of employers toward $c$~\eqref{eq:employer-demand} likely to exceed the demands of $c$ toward employers~\eqref{eq:cand-demand}.
In this case, transfers $\tau_{c,\cdot}$ that the candidate $c$ requests to employers increase to adjust the excess demands in the market.
On the other hand, a popular employer $j$ with high $p_{c,j}$s tends to have excess demands from many candidates.
In this case, to adjust the excess demand toward $j$, the market would decrease $\tau_{\cdot,j}$ that the employer needs to pay when he matches.
After these adjustment processes, demands from both sides are settled in equilibrium.
Economists consider that the equilibrium is achieved in real-world matching markets such as job markets or marriage markets, and analyzes this equilibrium matching in TU matching model.
Online matching markets might also suffer from the excess demands toward a few popular users.
Therefore, we consider that the equilibrium in TU matching would be effective in designing a desirable reciprocal recommendation that takes into account two aspects discussed in the Introduction: the mutuality of preferences and the avoidance of concentration on a few popular users.

\citet{Galichon2022-dn} show that the equilibrium matching exists, can be computed as a solution to a convex optimization problem, and maximizes social gains under constraints.
However, solving this equilibrium matching, in general, is computationally difficult.
To derive tractable equations of equilibrium matching, we introduce the Gumbel specification~\cite{Choo2006-dd} of preference errors' distribution.

\begin{assumption}
\label{ass:gumbel}
Estimation errors of preference scores ($\epsilon_{c,j}$ for each $(c,j) \in \mC\times (\mJ\cup\{0\})$, and $\epsilon_{j,c}$ for each $(j,c) \in \mJ\times (\mC\cup\{0\})$) are identically and independently distributed from the Gumbel (type-I extreme value) distribution with location $0$ and scale $\beta > 0$.
\end{assumption}

Given Assumption~\ref{ass:gumbel}, we can show that $\mu^*_{c,j}$ for each $(c,j)\in \mJ\times \mC$ in equilibrium is
\begin{equation}\label{eq:equilibrium-matching-gumbel}
    \mu^*_{c,j} = \exp\left(\frac{p_{c,j}+p_{j,c}}{2\beta}\right)\sqrt{\mu_{c,0}^*}\sqrt{\mu_{0,j}^*}
\end{equation}
where $\mu_{c,0}^*, \mu_{0,j}^*$ is the stochastic demand of $c$/$j$ toward the outside option $0$.
In our method, we compute this equilibrium matching $\mu_{c,j}^*$ and generate the ranked list of employers in the descending order of $\mu_{c,\cdot}^*$ to each candidate $c$.
Unlike standard reciprocal recommendation methods that use aggregate functions, such as the product function discussed Section~\ref{subsec:existing-methods}, the equilibrium matching~\eqref{eq:equilibrium-matching-gumbel} is computed with not only the preference scores $p_{c,j}, p_{j,c}$ between the two users, but also with $\mu_{c,0}^*,\mu_{0,j}^*$ determined by the overall balance of the market.

\begin{algorithm*}[t]
    \caption{Solving Choo--Siow Equilibrium Matching via IPFP}\label{alg:ipfp}
    \SetKw{initialize}{initialize}
    \SetKw{return}{return}
    \KwIn{preference scores $(p_{c,j})_{(c,j)\in \mC\times \mJ}$ and  $(p_{j,c})_{(j,c)\in \mJ\times \mC}$, scale parameter $\beta > 0$, timesteps $T$}
    \KwOut{Equilibrium Matching $\mu^* = (\mu^*_{c,j})_{(c,j) \in \mC\times\mJ}$}
    \initialize{$A_c \leftarrow 1\ \forall c \in \mC$, $B_j \leftarrow 1\ \forall j \in \mJ$}\;
    \For{$t = 1,\dots T$}{
        $A_c \leftarrow \sqrt{1 + \left(\frac{1}{2}\sum_{j \in \mJ}\exp\left(\frac{p_{c,j}+p_{j,c}}{2\beta}\right)B_j\right)^2} - \frac{1}{2}\sum_{j \in \mJ}\exp\left(\frac{p_{c,j}+p_{j,c}}{2\beta}\right)B_j$ \ for each $c \in \mC$\;
        $B_j \leftarrow \sqrt{1 + \left(\frac{1}{2}\sum_{c \in \mC}\exp\left(\frac{p_{c,j} + p_{j,c}}{2\beta}\right)A_c\right)^2} - \frac{1}{2}\sum_{c \in \mC}\exp\left(\frac{p_{c,j} + p_{j,c}}{2\beta}\right)A_c$ \ for each $j \in \mJ$\;
    }
    $\mu_{c,j}^* \leftarrow \exp\left(\frac{p_{c,j} + p_{j,c}}{2\beta}\right) A_c B_j$ for each $(c,j) \in C\times\mJ$\;
    \return{$\mu^*$}\;
\end{algorithm*}

Now we can derive a system of equations to find the equilibrium matching.
Let $A_c \coloneqq \sqrt{\mu_{c,0}^*}$ and $B_j \coloneqq \sqrt{\mu_{0,j}^*}$. 
The equilibrium matching $\mu^*$ satisfies the following system of equations:
\begin{align}
    &A_c^2 + A_{c}\sum_{j \in \mJ}\exp\left(\frac{p_{c,j}+p_{j,c}}{2\beta}\right)B_j = 1 \qquad \text{for each } c \in \mC, \label{eq:ipfp-1}\\
    &B_j^2 + B_j\sum_{c \in \mC}\exp\left(\frac{p_{c,j}+p_{j,c}}{2\beta}\right)A_c = 1 \qquad \text{for each } j \in \mJ,\label{eq:ipfp-2}\\
    &\mu_{c,j}^* = \exp\left(\frac{p_{c,j} + p_{j,c}}{2\beta}\right)A_c B_j \qquad \text{for each } (c,j) \in \mC \times \mJ, \label{eq:ipfp-3}
\end{align}
The derivation of these equations is shown in Appendix~\ref{subsec:derivate-equations}.

These equations can be solved in several methods.
In our experiments discussed later, we solve the equations via IPFP (Iterative Proportional Fitting Procedure)~\cite{decker2013unique,Galichon2022-dn}, which is summarized as Algorithm~\ref{alg:ipfp}.

\subsection{Computational efficiency in inference time}\label{subsec:inference}
As mentioned above, the existing method requires precomputing Birkhoff--von-Neumann decomposition~\cite{birkhoff1946three} of a doubly stochastic matrix for each job candidate to efficiently sample a deterministic ranking in inference time.
However, it leads to a prohibitively large computational cost, being infeasible in realistic applications.
The space cost for the precomputed set of top-$K$ deterministic rankings for each job candidate may also be impractical.    
By contrast, our method enables efficient real-time retrieval via vector search algorithms~\cite{shrivastava2014asymmetric} when employing two-tower/dot-product models~\cite{rendle2020neural}, such as matrix factorization~\cite{hu2008collaborative,koren2009matrix}, for preference scores; this is an essential property for large-scale applications (e.g., \cite{covington2016deep}).
Assume that preference scores can be expressed in the following dot-product models,
\begin{align}
  p_{c,j} = \langle \phi_1(c), \psi_1(j)\rangle, \quad p_{j,c} = \langle \phi_2(c), \psi_2(j)\rangle. \notag
\end{align}
Here, $\phi_1$ and $\phi_2$ are the $d$-dimensional feature mappings for job candidates in the two preference models,
while $\psi_1$ and $\psi_2$ are those for job posts.
Recall here that our method predicts a ranked list of $\mJ$ for a candidate $c \in \mC$ according to the order of $\{\mu^*_{c,j}\}_{j \in \mJ}$, and hence it is equivalent to use $\{\log(\mu^*_{c,j})\}_{j \in \mJ}$ to generate ranked lists due to the monotonicity of $\log(\cdot)$.
By using Equation~\eqref{eq:equilibrium-matching-gumbel},
this equivalent ranking score can be reformulated as follows:
\begin{align*}
  &\log(\mu^*_{c,j})\\
  =& \log\left(\exp\left(\frac{p_{c,j} + p_{j,c}}{2\beta}\right)\sqrt{\mu^*_{c,0}}\sqrt{\mu_{0,j}^*}\right) \\
  =& \frac{p_{c,j} + p_{j,c}}{2\beta} + \frac{1}{2}\log(\mu^*_{c,0}) + \frac{1}{2}\log(\mu_{0,j}^*) \\
  =& \frac{1}{2\beta}\left(\langle \phi_1(c), \psi_1(j)\rangle + \langle \phi_2(c), \psi_2(j)\rangle\right) + \frac{1}{2}\log(\mu^*_{c,0}) + \frac{1}{2}\log(\mu_{0,j}^*) \\
  =& \frac{1}{2\beta}\left\langle\left[\phi_1(c), \phi_2(c)\right], \left[\psi_1(j), \psi_2(j)\right]\right\rangle + \frac{1}{2}\log(\mu^*_{c,0}) + \frac{1}{2}\log(\mu_{0,j}^*) \\
  =& \frac{1}{2\beta}\left\langle \left[\phi_1(c), \phi_2(c), \beta\log(\mu^*_{c,0}), 1 \right], \left[\psi_1(j), \psi_2(j), 1, \beta\log(\mu_{0,j}^*)\right]\right\rangle,
\end{align*}
where $[\mathbf{u}, \mathbf{v}]$ is the concatenation of two vectors $\mathbf{u}$ and $\mathbf{v}$.
Therefore, our method maintains the desirable structure in dot-product models of preference scores by using the following $(2d + 2)$-dimensional feature mappings,
\begin{align*}
  &\phi(c) = \left[\phi_1(c), \phi_2(c), \beta\log(\mu^*_{c,0}), 1 \right],\\
  &\psi(j) = \left[\psi_1(j), \psi_2(j), 1, \beta\log(\mu_{0,j}^*)\right].
\end{align*}
Note here that we omitted the positive constant $1/2\beta$ as it does not affect ranking predictions.

\section{Experiments with synthetic data}
\label{sec:synthetic}
In this section, we report experimental results that compare our proposed method with other methods on synthetic datasets\footnote{The codes are available at \url{https://github.com/CyberAgentAILab/tu-matching-recommendation}.}.
The experiment setup in this section strictly follows \citet{Su2022-hx}.

\subsection{Setup}
\paragraph{Data generation}
We generated synthetic data according to the setup of \citet{Su2022-hx}.
The number of employers is $|\mJ| = n$ and that of job candidates is $|\mC| = 1.5n$ where $n$ is the market size parameter in $\{50, 100, 200, 500\}$.

To simulate various scenarios, we divide each unilateral preference score into two terms: $p_{c,j} \coloneqq \lambda\overline{p}_j + (1-\lambda) \tilde{p}_{c,j}$ and $p_{j,c} \coloneqq \lambda \overline{p}_c + (1-\lambda)\tilde{p}_{j,c}$ for each $(c,j) \in \mC\times \mJ$, where the first term is the overall popularity of the user, and $\lambda \in [0,1]$ is its weight representing the degree of \emph{crowding}.
Here, employers and job candidates are indexed as $j_1,j_2,\dots,j_{|\mJ|},c_1,c_2,\dots,c_{|\mC|}$,
and we ensure that employers/candidates with low indexes are generally popular with all candidates/employers, i.e., $\overline{p}_{j_k} = 1 - \frac{k-1}{|\mJ|-1}$ for each $k = 1,\dots,|\mJ|$, and $\overline{p}_{c_k} = 1 - \frac{k-1}{|\mC|-1}$ for each $k = 1, \dots, |\mC|$.
The second term is the individual preference of each user.
We independently and identically draw $\tilde{p}_{c,j}, \tilde{p}_{j,c}$ from uniform distribution $U[0,1]$.
The crowding level in the market is controlled by the crowding parameter $\lambda \in \{0.0, 0.25, 0.5, 0.75, 1.0\}$.
We consider three types of examination functions $v$, i.e., (i) \textbf{``inv''}: $v(k) = 1/k$, (ii) \textbf{``exp''}: $v(k) = 1/\exp(k-1)$, and (iii) \textbf{``log''}: $v(k) = 1/\log(k+1)$.
Unless mentioned, we use $n = 100$, $\lambda = 0.5$, and $v(k) = 1/k$ (i.e., ``inv'') for all users in the experiments.

\paragraph{Metrics}
We evaluate our method and baselines in terms of the expected total number of matches $\sw(\pi)$, which is defined by Equation~\eqref{eq:sw}.
We estimate it by $10{,}000$ Monte Carlo simulations of the interaction processes in the matching market, which is described in Section~\ref{subsec:framework}.
For each setting, we repeatedly run this experiment 10 times and report the average results.

\paragraph{Methods compared}
We compare our TU method with three baselines, \emph{Naive}, \emph{Reciprocal}, and \emph{SW}~\cite{Su2022-hx} methods, which are discussed in Section~\ref{subsec:existing-methods}.
We set learning rate $\eta_t = 0.2$ and timesteps $T=50$ for the \emph{SW} method following \citet{Su2022-hx}.
For our \emph{TU} method, we use $\beta = 1.0$.
We also tested other values of $\beta$, but the results are similar to the case of $\beta = 1.0$.
The detail for effects of $\beta$ is discussed in Appendix~\ref{subsec:effects-beta}.
In all cases, we confirmed that the IPFP algorithm converges in less than $50$ steps\footnote{We decide that the procedure has converged if both updates of all $A_c, B_j$ in one step and errors of the constraints \eqref{eq:constraint1}--\eqref{eq:constraint2} have become less than 1e-9.}.

\subsection{Results}

\begin{figure*}[t]
    \begin{minipage}[t]{\linewidth}
    \centering
    \includegraphics[width=\linewidth]{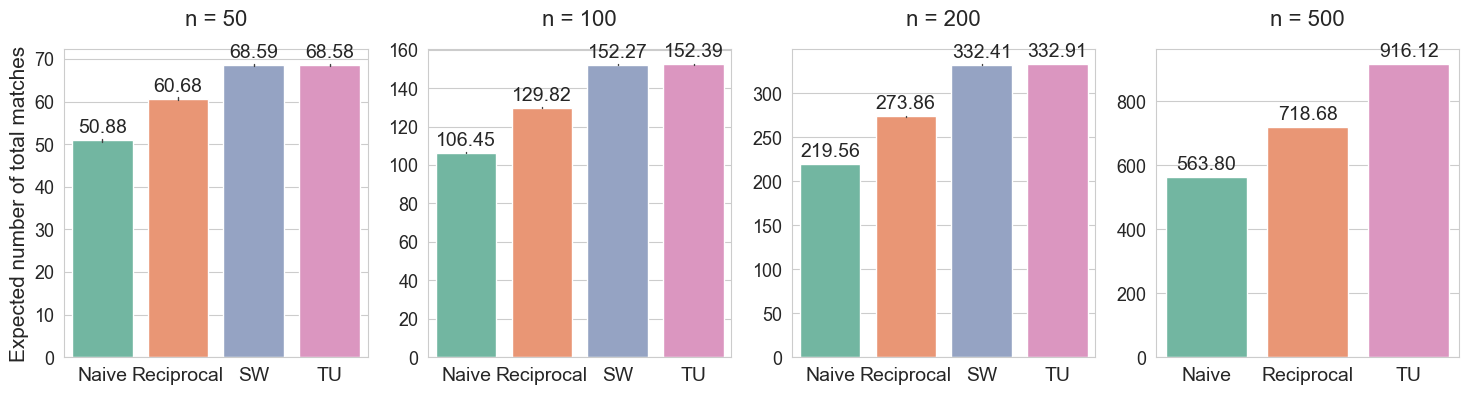}
    \caption{Results of synthetic data experiments with various market sizes $n$. The crowding parameter is $\lambda = 0.5$, and the examination function is $v(k) = 1/k$ (inv). Since the variation in the 10 experiments is small and the standard errors are on the order of 1e-1, the error bars are extremely small or invisible. In the case of $n = 500$, \emph{SW} method can not be computed, and thus we only report the result of \emph{Naive}, \emph{Reciprocal}, and \emph{TU} methods.}
    \label{fig:fig-n}
    \end{minipage}\\
    \begin{minipage}[t]{\linewidth}
    \centering
    \includegraphics[width=\linewidth]{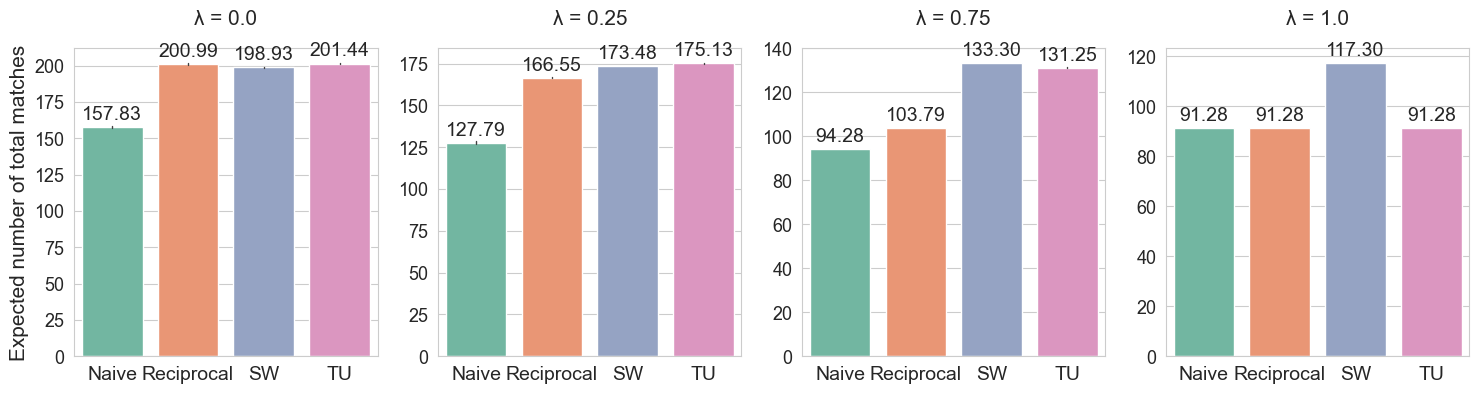}
    \caption{Results of synthetic data experiments with various levels of crowding parameters $\lambda$. Market size is $n = 100$, and the examination function is $v(k) = 1/k$ (inv). The case of $\lambda = 0.5$ is omitted in this figure because it is the case of $n = 100$ in Fig.~\ref{fig:fig-n}.}
    \label{fig:fig-lambda}
    \end{minipage}\\
    \begin{minipage}[t]{\linewidth}
    \centering
    \includegraphics[width=\linewidth]{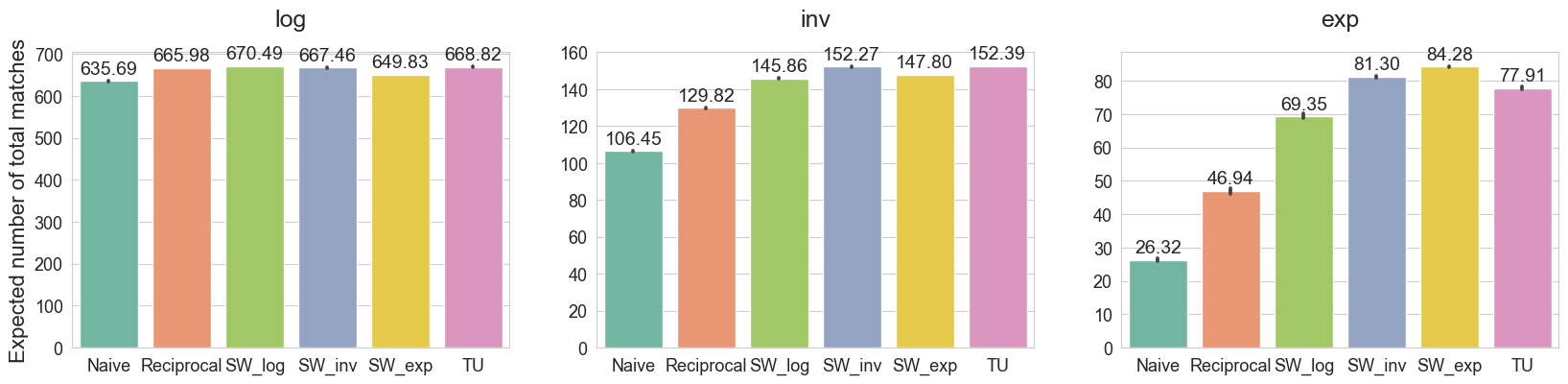}
    \caption{Results of synthetic data experiments with various examination functions $v(k) = 1/\log(k+1)$ (log), $1/k$ (inv), and $1/\exp(k-1)$. We also test \emph{SW} methods with different examination functions. Market size is $n = 100$, and the crowding parameter is $\lambda = 0.5$.}
    \label{fig:fig-v}
    \end{minipage}
\end{figure*}

\paragraph{Market sizes}
We test for various market sizes $n = 50, 100, 200$, and $500$.
The results are reported in Fig.~\ref{fig:fig-n}.
Although the total number of matches in all methods increases when the market size $n$ grows, the comparison results are almost unaffected by the market size.
\emph{TU} outperforms \emph{Naive} and \emph{Reciprocal} methods, and performs as well as \emph{SW} method.
We note that \emph{SW} cannot be computed for the case of $n = 500$, and hence we do not report its result for this case, which corresponds to the rightmost figure.
In addition, since the variation in the 10 experiments is very small and the standard errors are on the order of 1e-1, the error bars are extremely small or invisible in Fig~\ref{fig:fig-n}, which is also the case in Fig.~\ref{fig:fig-lambda} and Fig.~\ref{fig:fig-v}.

\paragraph{Crowding}
We also test for various levels of crowding $\lambda$ $=$ $0.0$, $0.25$, $0.5$, $0.75$, and $1.0$.
The results are shown in Fig.~\ref{fig:fig-lambda}.
The case of $\lambda = 0.5$ equals the case of $n = 100$ in Fig.~\ref{fig:fig-n}, and thus it is omitted in Fig.~\ref{fig:fig-lambda}.
With lower crowding parameter $\lambda = 0.0, 0.25$, which means that users' preferences vary and there are few inequalities of popularity among users, \emph{TU} is as well as \emph{Reciprocal} and \emph{SW}.
On the other hand, in the case of $\lambda$, which means that there are huge inequalities of popularity among users, \emph{TU} performs worse than \emph{SW}.
In the middle level of crowding $\lambda = 0.5, 0.75$, \emph{TU} outperforms \emph{Naive} and \emph{Reciprocal} while it performs as well as \emph{SW}.

\paragraph{Examination functions}
Finally, we test three types of examination functions (log, inv, and exp) used in the Monte Carlo simulations in the testing phase.
In addition, to assess the effect of misspecification of examination functions used in \emph{SW} discussed in Section~\ref{subsec:existing-methods}, we consider the variants of \emph{SW} with different examination functions in the training phase, namely, \emph{SW\_log}, \emph{SW\_inv}, and \emph{SW\_exp}, which use log, inv, and exp examinations in the objective of \emph{SW}, respectively.
Market size is $n = 100$, and the crowding parameter is $\lambda = 0.5$.
The results are reported in Fig.~\ref{fig:fig-v}.
In each case, the \emph{SW} variants with misspecified examination functions (e.g., \emph{SW\_inv} and \emph{SW\_exp} in the log case) perform worse than the \emph{SW} with correct examination functions.
Specifically, \emph{SW\_exp} underperforms even \emph{Reciprocal} method in the log case, and \emph{SW\_log} in the exp case performs 17.7 \% worse than \emph{SW\_exp}.
By contrast, \emph{TU} is competitive to \emph{SW} with correct examinations in log and inv cases and slightly deteriorates in the exp case.
In most cases, \emph{TU} outperforms \emph{SW} with misspecified examination functions, whereas it slightly underperforms \emph{SW\_inv} in the exp case.

\paragraph{Summary}

In summary, we show in these synthetic experiments that:
\begin{enumerate}
    \item \emph{TU} method performs as well as \emph{SW} method in many cases.
    \item It works well even in relatively large datasets for which \emph{SW} does not work, and substantially outperforms the conventional \emph{Naive} and \emph{Reciprocal} methods.
    \item It outperforms \emph{SW} methods with misspecified examination functions in many cases, except that \emph{SW}\_inv in the exp case; however, the performance gap between \emph{SW}\_inv and \emph{TU} is not substantial.
\end{enumerate}
In real-world platforms, it is often difficult to specify the examination function for each user accurately, and \emph{SW} is also infeasible even in the case with $n=500$.
Therefore, our TU method would be effective in practice.

\section{Experiments with real-world data}
\label{sec:semi-synthetic}

We conducted an experiment using real-world data collected from an online dating platform.
This service is a Japanese online dating platform with millions of cumulative members, where men are recommended to women and vice versa.
Users who receive recommendations choose ``like'' if they prefer the other person and ``dislike'' if they do not.
Furthermore, users who receive ``like'' respond with ``thank you'' if they like the other person and ``sorry'' if they do not. Only when a ``thank you'' is received does the match become established.
Then they will be able to chat.

\subsection{Setup}
\paragraph{Data preprocessing}
We explain how to create moderate-sized data using our raw, large data.
We first extracted two datasets, one consisting of $1{,}000$ sampled men and women (denoted $1{,}000 \times 1{,}000$ dataset) and the other consisting of $200$ sampled men and women (denoted $200 \times 200$ dataset).
Our preliminary experiments revealed that the baseline method of \citet{Su2022-hx} did not work on the $1{,}000 \times 1{,}000$ dataset due to its space and time complexity, whereas Monte Carlo simulations and our proposed method work.
Hence, we used the additional $200 \times 200$ dataset for comparison between our algorithm and the baselines.

Since random sampling may yield too sparse data to train MF-based preference models, we performed the following two steps to extract a subset of users that induces a dense component:
In the first step, we filtered users by their residence.
The intuition behind this filtering is that users are more likely to prefer each other if they are living in the same (or nearby) residence.
In the second step, we applied a $k$-core decomposition algorithm to the remaining data.
For a graph $G$, a \emph{$k$-core} is any maximal subgraph of $G$ such that every vertex has degree at least $k$~\cite{seidman1983network}\footnote{
In our dataset, each vertex corresponds to a user, and each edge represents an interaction between a pair of uses.}.
Therefore, a $k$-core (with large $k$) is able to find a dense part of $G$.
It should be noted that there are two directions of interaction due to the service's specifications:
(1) male-proactive scenario: men receiving a ``thank you'' for their ``like'' from women, and
(2) female-proactive scenario: women receiving a ``thank you'' for their ``like'' from men.
For pairs with unobserved evaluation data, we used the Alternating Least Squares (ALS) method to complete the data.

\paragraph{Experimental process}
To assess the performance of our method with real-world data, we conducted the Monte Carlo simulations as well as synthetic data simulations discussed in Section~\ref{sec:synthetic}.
Using the estimated preferences from real-world data discussed above, we ran \emph{Naive}, \emph{Reciprocal}, \emph{TU}, and \emph{SW} (only for the $200\times 200$ dataset) methods.
We used constant learning rate $\eta_t = 0.2$ and timesteps $T=50$ for the learning process of \emph{SW}, and $\beta = 1.0$ for \emph{TU}, similarly to the synthetic data experiments.
We used the inv examination function $v(k)=1/k$ for both the Monte Carlo simulations and the objective function of \emph{SW}.
We estimated the expected total number of  matches by running Monte Carlo simulations $10{,}000$ times for $200\times 200$ datasets, and $1{,}000$ times for $1{,}000\times 1{,}000$ datasets.

\subsection{Results}

\begin{figure*}[t]
    \centering
    \includegraphics[width=\linewidth]{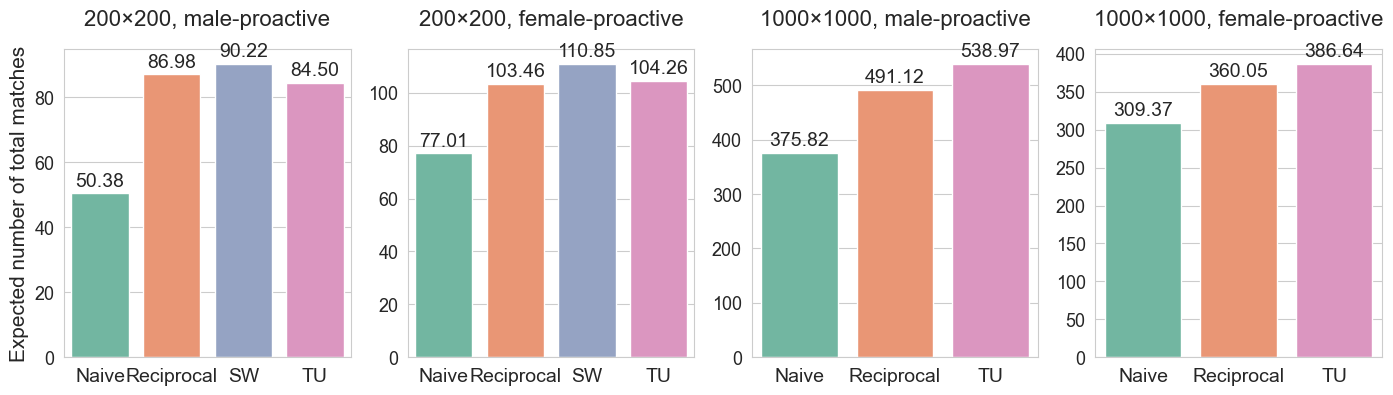}
    \caption{Results of experiments with real-world online dating data. Since \emph{SW} cannot be computed in $1{,}000 \times 1{,}000$ case, we omit the result of \emph{SW} in such cases.}
    \label{fig:fig-semi}
\end{figure*}

The results are reported in Fig.~\ref{fig:fig-semi}.
\emph{TU} method performs slightly worse than \emph{SW}, and as well as \emph{Reciprocal} in $200\times 200$ datasets for both male-proctive and female-proactive case.
We note that the inv examination function is used both for the objective function of \emph{SW} and the Monte Carlo simulation process and the issue of misspecification of $v$ is not considered in this simulation, and thus it is just an infeasible skyline method.
In $1{,}000\times 1{,}000$ datasets, \emph{TU} performs clearly better than \emph{Naive} and \emph{Reciprocal} methods.

As in the synthetic experiments, we show in these experiments with real-world data that the \emph{TU} method is also effective for realistic situations.
These results suggest that our \emph{TU} method would be beneficial in practice for real-world online dating or job posting platforms with more than thousands of users.

\section{Conclusions and Future works}
In this paper, we introduce the TU matching-based reciprocal recommender system that does not rely on the specific form of examination functions, and works in a relatively large datasets for which one existing method does not work.
We evaluate the performance of our method in experiments with synthetic and real-world data from an online dating platform.
Our \emph{TU} method outperforms the conventional \emph{Naive} and \emph{Reciprocal} methods, and performs at least as well as the \emph{SW} method.
These experiments suggest that the \emph{TU} method can be superior to \emph{SW} in a realistic situation where \emph{SW} uses a possibly misspecified examination function.

There are several considerable future directions in the field of reciprocal recommendation in matching markets.
First, in this paper, whereas we evaluate our \emph{TU} method only in experiments that run Monte Carlo simulations that simulate the matching markets,
\emph{evaluation in online A/B experiments} in real-world platforms remains for future works.
Moreover, it is also important to develop offline/off-policy evaluation methods in reciprocal recommendation in matching markets.

Next, \emph{more scalable ranking methods in RRSs} would also be a future direction.
Although our \emph{TU} method is more efficient in terms of time and space complexity than the \citet{Su2022-hx}'s method, it still costs the order of computational complexity $O(|\mJ||\mC|)$ for each timestep in Algorithm.~\ref{alg:ipfp}, and we only conduct experiments with $1{,}000\times 1{,}000$ datasets at most.
Since there are real-world matching platforms with tens or hundreds of thousands of users, a more efficient implementation of our \emph{TU} method or other algorithms would be needed for such large-scale matching platforms.

Another future direction is \emph{applications of other matching algorithms}.
Besides TU matching~\cite{shapley1971assignment,Choo2006-dd,Galichon2022-dn}, many other matching algorithms  have been developed in the field of matching with non-transferable utility (NTU matching), such as the differed acceptance (DA) algorithm of \citet{gale1962college}.
Only a few works have developed a reciprocal recommender system based on matching algorithms like DA algorithm~\cite{saini2019privatejobmatch,eskandanian2020using,pizzato2011stochastic}.
Especially, there seems to be possible interaction between the issues of fairness in the reciprocal recommendation and the matching with constraints.
In the literature of matching theory, various algorithms with complex constraints regarding fairness issues have been recently developed, such as NTU matching with affirmative actions~\cite{abdulkadirouglu2005college,hafalir2013effective} and with regional constraints~\cite{kamada2015efficient,aziz2020matching}, and TU matching with regional constraints~\cite{jalota2022matching,matsushita2022regulating}.
These matching algorithms with constraints could be beneficial to improve the fairness of reciprocal recommendations.

Finally, the issues of \emph{incentive and strategic behavior in reciprocal recommendation} would also be an important future direction.
In reciprocal recommendations, the problem could be whether users act according to their true preferences.
For example, a job seeker may not apply for a job slot in which she is strongly interested because the position seems to be popular among job seekers and highly competitive.
Thus, the action history obtained in reciprocal recommendations may not reflect the true preferences of users.
Furthermore, consider the situation where the reciprocal recommendation method other than the naive method that recommends naively according to the estimated preferences of proactive users is used.
Users may have incentives to manipulate their actions and reveal false preferences because the recommendation systems do not simply suggest users that they seem to be interested in.
The issue of strategic behavior has been extensively analyzed in matching theory, and many ``strategy-proof'' matching algorithms that incentivize users to reveal their true preferences have been developed~\cite{roth1989college,sonmez1994strategy}.
Designing RRSs that incentivize users to act according to their true preferences based on strategy-proof matching algorithms could be an interesting future direction.

\appendix
\section{Appendix}

\begin{table*}[t]
    \centering
    \caption{The performance and the number of iterations needed for convergence of the TU method with different $\beta$.}
    \label{tab:beta}
    \begin{tabular}{cccccccccc}
    \toprule
        Method& \multirow{2}{*}{Naive} & \multirow{2}{*}{Reciprocal} & \multirow{2}{*}{SW} & \multicolumn{6}{c}{TU} \\
        $\beta$ & & & & $0.1$ & $0.5$ & $1.0$ & $2.0$ & $5.0$ & $10.0$ \\ \hline
        \multirow{2}{*}{\# of matches} & 106.450 & 129.824 & 152.269 & 152.318 & 152.365 & 152.389 & 152.460 & 152.722 & 153.089\\
        & (0.176) & (0.178) & (0.101) & (0.104) & (0.104) & (0.105) & (0.096) & (0.102) & (0.095) \\
        \# of iterations & & & & 39 & 39 & 40 & 42 & 48-49 & >100{,}000 \\ 
    \bottomrule
    \end{tabular}
\end{table*}

\begin{figure*}[t]
    \centering
    \includegraphics[width=\linewidth]{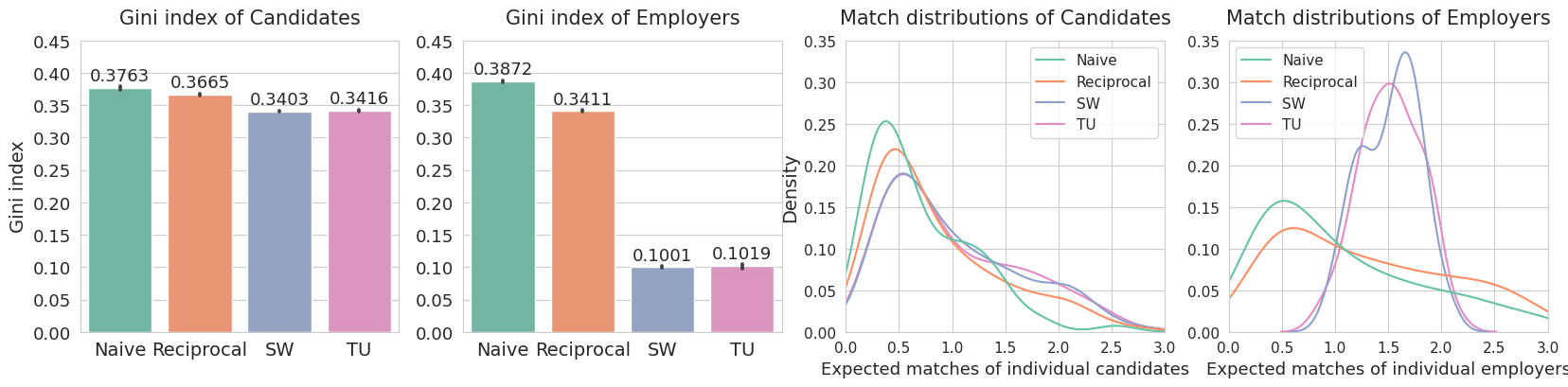}
    \caption{The two left graphs shows the Gini index of matches among candidates and employers in the synthetic simulations with $n= 100$, $\lambda = 0.5$, and $v$ is the inv case. The two right graphs are the distributions of the expected number of matches of individual candidates and employers in one representative simulation.}
    \label{fig:fig-fairness}
\end{figure*}

\subsection{Derivation of Equations in IPFP}
\label{subsec:derivate-equations}

Given Assumption~\ref{ass:gumbel} and transfers $\tau$, the stochastic demands of candidates~\eqref{eq:cand-demand} and employers~\eqref{eq:employer-demand} can be written as
\begin{align}
    \mu_{c,j} &= \bP_{\epsilon_{c,\cdot}}\left( j \in \argmax_{j' \in \mJ\cup\{0\}}\left(p_{c,j'} + \epsilon_{c,j'} + \tau_{c,j'}\right)\right) \notag\\
    &= \frac{\exp\left((p_{c,j} +\tau_{c,j})/\beta\right)}{\sum_{j' \in \mJ\cup\{0\}}\exp\left( (p_{c,j'} + \tau_{c,j'})/\beta \right)} \label{eq:app_mu_c_j}
\end{align}
for each $(c,j) \in \mC \times (\mJ\cup\{0\})$ (here we note that $p_{c,0} = \tau_{c,0} = 0$) and
\begin{align}
    \mu_{j,c} &= \bP_{\epsilon_{j,\cdot}}\left( c \in \argmax_{c' \in \mC\cup\{0\}}\left(p_{j,c'} + \epsilon_{j,c'} - \tau_{c',j}\right) \right)\notag\\
    &= \frac{\exp\left((p_{j,c} - \tau_{c,j})/\beta\right)}{\sum_{c'\in\mC\cup\{0\}}\exp\left((p_{j,c'} - \tau_{c',j})/\beta\right)} \label{eq:app_mu_j_c}
\end{align}
for each $(j,c) \in \mJ \times (\mC\cup\{0\})$ (also note that $p_{j,0} = \tau_{0,j} = 0$).
In the equilibrium matching $(\mu^*, \tau^*)$, we can derive
\begin{align}
    \notag
    \mu^{*}_{c,j} =& \sqrt{\mu_{c,j}}\sqrt{\mu_{j,c}}\\
    =& \sqrt{\frac{\exp\left((p_{c,j} +\tau^*_{c,j})/\beta\right)}{\sum_{j' \in \mJ\cup\{0\}}\exp\left( (p_{c,j'} + \tau^*_{c,j'})/\beta \right)}} \notag\\
    =& \exp\left(\frac{p_{c,j} + \tau^*_{c,j}}{2\beta}\right)\exp\left(\frac{p_{j,c}-\tau^*_{c,j}}{2\beta}\right) \notag\\
    =& \exp\left(\frac{p_{c,j} + p_{j,c}}{2\beta}\right)\sqrt{\mu^*_{c,0}}\sqrt{\mu_{0,j}^*},
    \label{eq:mu^*_c,j}
\end{align}
where the first equality holds because $\mu^*_{c,j} = \mu_{c,j} = \mu_{j,c}$~\eqref{eq:cand-demand}--\eqref{eq:employer-demand} in the equilibrium and the second equality holds since ~\eqref{eq:app_mu_c_j}--\eqref{eq:app_mu_j_c}.
Let $A_c \coloneqq \sqrt{\mu^*_{c,0}}$ and $B_j \coloneqq \sqrt{\mu^*_{0,j}}$.
Combining \eqref{eq:mu^*_c,j} with constraints \eqref{eq:constraint1}--\eqref{eq:constraint2}, we have the equations~\eqref{eq:ipfp-1}--\eqref{eq:ipfp-3}.

\subsection{Effects of $\beta$}
\label{subsec:effects-beta}

The only hyperparameter that needs to be tuned in the Algorithm~\ref{alg:ipfp}, other than timesteps or stopping criterion, is the scale parameter $\beta$ for the Gumbel distribution of preference errors in Assumption~\ref{ass:gumbel}.
To evaluate the effects of $\beta$ on the performance of the \emph{TU} method, we test $\beta = 0.1, 0.5, 1.0, 2.0, 5.0$ and $10.0$ for the case of $n = 100$, $\lambda = 0.5$, and $v(k) = 1/k$ (inv) in the synthetic data experiments.
The results are reported in Table~\ref{tab:beta}.
Although larger $\beta$ results in slightly better performances in terms of the expected total number of  matches, it seems that values of $\beta$ do not have a large impact on performances.
However, the number of iterations needed for convergence increases when $\beta$ increases.
Especially, the procedure did not converge even after $100{,}000$ iterations for $\beta = 10.0$, where we report the performance of $\mu$ after $100{,}000$ iterations.

\subsection{Discussion on Fairness}
\label{subsec:fairness}

In matching markets such as job posting and online dating platforms, not only is the total number of matches important but also the fairness between users. 
In addition, it may be effective even for the total number of matches to distribute matches fairly among users rather than to concentrate matches to a few popular users.
The two left graphs in Fig.~\ref{fig:fig-fairness} show the Gini index of matches among candidates and employers in synthetic experiments in the case of $n = 100$, $\lambda = 0.5$, and the inv examination function.
The mean of 10 simulations is reported.
A lower Gini coefficient indicates a more equal distribution.
In the two right graphs, the distributions of matches among candidates and employers in one representative simulation are shown.
In both graphs, we can see that \emph{TU} is as fair as \emph{SW}, and fairer than \emph{Naive} and \emph{Reciprocal} methods.
Especially, the equality of matches among employers (reactive side) substantially improves in \emph{TU} and \emph{SW} methods from \emph{Naive} and \emph{Reciprocal} methods.

\bibliographystyle{ACM-Reference-Format}
\bibliography{references}

\end{document}